\documentclass[aps,prb,reprint,superscriptaddress,showpacs,showkeys,floatfix]{revtex4-1}

\usepackage{graphicx,amsmath,xspace}
\usepackage[per-mode=symbol]{siunitx}
\DeclareSIUnit\atom{at}
\newcommand{\ZrAlN}{Zr$_{1-x}$Al$_{x}$N\xspace}

\begin{document}

\title{Macroscopic Elastic Properties of Textured ZrN--AlN Polycrystalline Aggregates: From Ab initio Calculations to Grain-Scale Interactions}

\author{D. Holec}
  \email{david.holec@unileoben.ac.at}
  \affiliation{Department of Physical Metallurgy and Materials Testing, Montanuniversit\"at Leoben, A-8700 Leoben, Austria}
\author{F. Tasn\'adi}
  \affiliation{Department of Physics, Chemistry and Biology (IFM), Link\"oping University, SE-581 83 Link\"oping, Sweden}  
\author{P. Wagner}
  \affiliation{Department of Physical Metallurgy and Materials Testing, Montanuniversit\"at Leoben, A-8700 Leoben, Austria}
\author{M. Fri\'ak}
  \affiliation{Max-Planck-Institut f\"ur Eisenforschung GmbH, D-40237 D\"usseldorf, Germany}
  \affiliation{Institute of Physics of Materials of the Academy of Sciences of the Czech republic, v.v.i., CZ-61662 Brno, Czech Republic}
\author{J. Neugebauer}
  \affiliation{Max-Planck-Institut f\"ur Eisenforschung GmbH, D-40237 D\"usseldorf, Germany}  
\author{P.H. Mayrhofer}
  \affiliation{Institute of Materials Science and Technology, Vienna University of Technology, Vienna, A-1140, Austria}
\author{J. Keckes}
  \affiliation{Erich Schmid Institute of Materials Science, Austrian Academy of Sciences, A-8700 Leoben, Austria}
  \affiliation{Department of Materials Physics, Montanuniversit\"at Leoben, A-8700 Leoben, Austria}
\date{\today}

\begin{abstract}
  Despite the fast development of computational materials modelling, theoretical description of macroscopic elastic properties of textured polycrystalline aggregates starting from basic principles remains a challenging task. In this study we use a supercell-based approach to obtain the elastic properties of random solid solution cubic \ZrAlN system as a function of the metallic sublattice composition and texture descriptors. The employed special quasi-random structures are optimised not only with respect to short range order parameters, but also to make the three cubic directions $[1\,0\,0]$, $[0\,1\,0]$, and $[0\,0\,1]$ as similar as possible.  In this way, only a small spread of elastic constants tensor components is achieved and an optimum trade-off between modelling of chemical disorder and computational limits regarding the supercell size and calculational time is proposed. The single crystal elastic constants are shown to vary smoothly with composition, yielding $x\approx0.5$ an alloy constitution with an almost isotropic response. Consequently, polycrystals with this composition are suggested to have Young's modulus independent on the actual microstructure. This is indeed confirmed by explicit calculations of polycrystal elastic properties, both within the isotropic aggregate limit, as well as with fibre textures with various orientations and sharpness. It turns out, that for low AlN mole fractions, the spread of the possible Young's moduli data caused by the texture variation can be larger than $\SI{100}{\giga\pascal}$. Consequently, our discussion of Young's modulus data of cubic \ZrAlN contains also the evaluation of the texture typical for thin films.
\end{abstract}

%\keywords{}

\maketitle

\section{Introduction}

Quantum mechanical calculations using Density Functional Theory (DFT) of structural as well as elastic properties of materials have become a standard tool in modern computational material science. Recently, also the alloying trends have been heavily investigated, which in the area of hard protective coatings addressed predominantly issues related to the phase stability (see e.g. Refs. \onlinecite{Mayrhofer2006a, Alling2007, Sheng2008, Matenoglou2009, Sangiovanni2011, Holec2013a}). This has been possible due to the increased computational power, and the development of theories for treating random solid solutions. These include effective potential methods\cite{Gonis2000-xb} (e.g., coherent potential approximation or virtual coherent approximation), cluster methods \cite{Sanchez2010-qb} (e.g., cluster expansion method), or supercell-based approaches, such a quasi-random structures (SQSs) \cite{Wei1990a} technique employed in this paper.

While the bulk modulus is relatively easy to obtain from the Birch-Murnaghan equation of state\cite{Birch1947} as used during the structure optimisation, the full tensor of elastic constants, $C_{ij}$, requires additional calculations. The two common methods to calculate $C_{ij}$ from the first principles are the total energy method and the stress-strain method. The latter relies on availability of stress tensor and uses Hooke's law to evaluate $C_{ij}$ directly. On the other hand, the total energy method assigns an energy difference between a ground and a deformed state to the strain energy. This is a function of applied strain and a specific combination of the elastic constants. The advantage of this method is that the total energy is always available from \textit{ab initio} calculations, and it furthermore allows for estimation of higher order elastic constants \cite{Holec2012}. The disadvantage is that it usually takes more CPU resources than the stress-strain method as more deformation modes need to be applied. It has been also recently proposed that the stress-strain method is a more robust technique \cite{Caro2013}.

When it comes to the elastic constants of materials without any long-ranged periodicity, the supercell approach faces an apparent problem: on one hand, the distribution of atoms on the lattice sites is required to be as random as possible to mimic solid solutions, hence often leading to supercells with only primitive symmetry (space group $P1$). On the other hand, the material is expected to exhibit certain symmetry based on its underlying lattice, for example the cubic symmetry of nitride coatings with the B1 (NaCl) structure. A combined \textit{ab initio} and molecular dynamics study \cite{VonPezold2010} has shown, that when the supercell is large enough, the differences between macroscopically equivalent directions or deformation modes (e.g., tension along the $x$, $y$, and $z$ direction in the cubic systems) vanish. Although this is promising, the idea is not in line with the original purpose of SQS which was to simulate random alloys with as small supercells as possible. \citet{Moakher2006} provided a rigorous mathematical theory on how to project a tensor of elastic constants with an arbitrary symmetry onto a tensor with a desired crystallographic symmetry. This has been applied to the cubic Ti$_{1-x}$Al$_x$N system \cite{Tasnadi2010a,Tasnadi2012} with a satisfactory agreement to available experimental data, however still requiring supercells with around 100 atoms and averaging over crystallographically equivalent directions.

In this work we investigate a possible trade-off between the randomness and the overall effective symmetry by introducing directionally-optimised SQSs (do-SQS, detailed description is given in Section~\ref{sec:supercell}) with the aim that the resulting tensor of elastic constants exhibits as small as possible deviations between the equivalent elastic constants. This in turn can lead to a significant reduction of computational resources by applying only a reduced set of deformations (similar to what is done to perfectly ordered and fully symmetric compounds, e.g. Ref.~\onlinecite{Holec2012}). The second part of this work is devoted to establishing the impact of a texture on the elastic constants of the polycrystalline aggregate. This is an important step towards a quantitative comparison of theoretical and experimental data, as well as theory-guided prediction of thin-film growth directions that provide extremal mechanical properties. Additional improvements towards modelling of real materials would be finite temperature effects and inclusion of grain boundaries, neither of which is addressed here.

To assess the performance of the here developed supercells, we have chosen cubic \ZrAlN system (NaCl prototype, $Fm\bar{3}m$ space group). It is an isovalent system with well investigated and widely used Ti$_{1-x}$Al$_x$N. Compared with TiN, ZrN has a lower coefficient of friction and has been suggested to have better oxidation resistance \cite{Chen2011,Sheng2008}. Additionally, calculated elastic constants of this system have not yet been published, and experimental values are only scarce.

\section{Methods}

\subsection{Supercells}\label{sec:supercell}

Warren-Cowley short-range order (SRO) parameters, $\alpha_j$, are commonly used to quantify randomness of an atom distribution on lattice sites. For binary alloys (or pseudo-binary, e.g. where the mixing happens only on one sublattice, as in the case of \ZrAlN), they are calculated as \cite{Muller2003}
\begin{equation}
  \alpha_j=1-\frac{N^j_{AB}}{x_Ax_BNM^j}\ ,
\end{equation}
where $x_A$ and $x_B$ ($x_A+x_B=1$) are the mole fractions of atoms $A$ and $B$, respectively, $N$ is the number of sites in the supercell, $M^j$ is the site coordination in the $j$th-neighbour distance, $d_j$, and $N^j_{AB}$ is the total number of $\{A,B\}$ pairs of atoms separated by the $d_j$ (number of $A$--$B$ bonds of length $d_j$). This definition implies that $\alpha_j>0$ and $\alpha_j<0$ correspond to tendency for clustering and ordering, respectively, while $\alpha_j=0$ describes an ideal statistically random alloy. When constructing special quasi-random structures (SQSs), the aim is to minimise values $|\alpha_j|$ for several first coordination shells (typically between $5$ and $7$).

\begin{figure}
  \includegraphics[width=0.8\columnwidth]{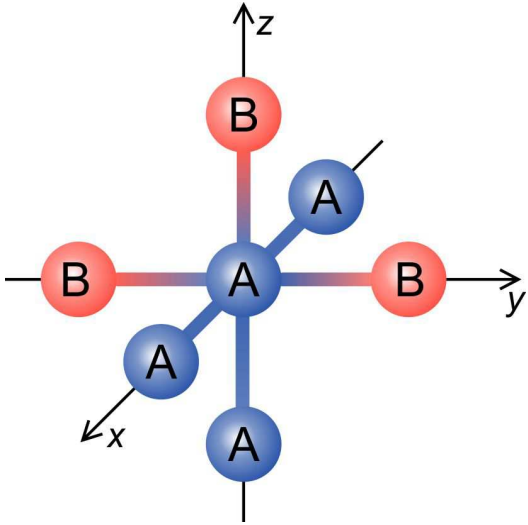}
  \caption{A schematic drawing of the do-SQS approach: the environment of the central atom A consists of 2 $A$--$A$ bonds in the $x$ direction, 2 $A$--$B$ bonds in the $y$ direction, and 1 $A$--$A$ and 1 $A$--$B$ in the $z$ direction. The same environment is described as 3 $A$--$A$ and 3 $A$--$B$ bonds within the SQS approach.}\label{fig:doSQS}
\end{figure}

\citet{Tasnadi2012} recently concluded that relatively large (around $100$ atoms and more) supercells are needed to accurately describe the elastic response of a cubic Ti$_{0.5}$Al$_{0.5}$N. Nevertheless, somewhat smaller cells with $64$ atoms and overall cubic shape do performed with an acceptable accuracy, too \cite{Tasnadi2012}. Moreover, we have applied a following additional constrain during the SQS generation: the number of bonds, $N_{AB}^j$, is divided into three subsets $N_{AB,x}^j$, $N_{AB,y}^j$, and $N_{AB,z}^j$, depending on which projection of the vector $\vec{AB}$ into $x$, $y$, and $z$ direction is the longest (Fig.~\ref{fig:doSQS}). Since the three directions $x$, $y$, and $z$ are crystallographically equivalent in the cubic systems, the projected SRO parameters are calculated as
\begin{equation}
  \alpha_{j,\xi}=1-\frac{N^j_{AB,\xi}}{\frac13x_Ax_BNM^j}\ ,\quad \xi=x,y,z\ .\label{eq:doSRO}
\end{equation}
This way, the number of $A$--$B$ bonds is optimised with respect to the three equivalent directions. We applied this requirement also to the $A$--$A$ and $B$--$B$ bonds. The resulting supercells, hereafter called \emph{directionally-optimised SQSs (do-SQSs)}, are summarised in Table~\ref{tab:supercells}. They were generated using a script which randomly distributes atoms $A$ and $B$ (considering a required chemical composition) on the (sub)lattice, hence providing a large ensemble of various atomic arrangements. For every one of them, projected SROs $\alpha_{j,\xi}$ up to $j=5$ were evaluated, and a supercell with $\alpha_{j,\xi}$ closest to 0, i.e. an ideal solid solution, was chosen. The projected SROs of the resulting supercells are listed in Table~\ref{tab:doSRO}. It is worth noting that the compositions $x=0.125$ ($x=0.875$) and $x=0.375$ ($=0.625$) are worse optimised than the other two $x=0.25$ ($x=0.75$) and $x=0.5$, a behaviour consistent with the analysis of standard SQSs reported in Ref.~\onlinecite{Alling2007a}. Finally, although the do-SQS method is developed here for high symmetry cubic systems, it can be applied also to other crystallographic classes by requesting the number of different bonds to be as similar as possible along equivalent directions.

\subsection{Elastic properties}

The single crystal elastic constants were obtained using the total energy method, discussed in detail in Ref.~\onlinecite{Holec2012}. The applied deformation matrices were
\begin{equation}
  A1=\begin{pmatrix} \delta & 0 & 0 \\ 0 & 0 & 0 \\ 0 & 0 & 0 \end{pmatrix},\, A2=\begin{pmatrix} \delta & 0 & 0 \\ 0 & \delta & 0 \\ 0 & 0 & 0 \end{pmatrix},\, A4=\begin{pmatrix} \delta & 0 & 0 \\ 0 & 0 & \delta \\ 0 & \delta & 0 \end{pmatrix}\ .
\end{equation}
The volumetric density of the total energy increase after application of such deformation matrix is, in the first approximation, a quadratic function of $\delta$
\begin{equation}
  U_A(\delta)={\cal A}\delta^2\ . \label{eq:quadratic}
\end{equation}
The coefficient $\cal A$ depends on the applied deformation matrix $A$. For example, for $A1$ it is equal to $\frac12C^{P1}_{11}$. By changing the position of the non-zero component $\delta$ in the matrix $A1$ to position $(2,2)$ and $(3,3)$, the coefficient ${\cal A}$ changes to $\frac12C^{P1}_{22}$ and $\frac12C^{P1}_{33}$, respectively. By fitting the $U_A(\delta)$ curves for various deformation matrices (coefficients ${\cal A}$ are listed in Table~\ref{tab:coeffsA}), a full tensors $C^{P1}_{ij}$ of single crystal elastic constants was obtained for every composition. In doing so, the off-diagonal components for rows $i=4,5,6$ and for columns $j=4,5,6$ in $C^{P1}_{ij}$ were fixed to 0.

\begin{table}
\caption{Quadratic coefficient ${\cal A}$ in the expansion of the strain energy (Eq.~\ref{eq:quadratic}) for various deformation matrices.}\label{tab:coeffsA}
\begin{ruledtabular}
\begin{tabular}{ccc}
  $A1$ & $(1,1)$ & $\frac12C^{P1}_{11}$\\
  $A1$ & $(2,2)$ & $\frac12C^{P1}_{22}$\\
  $A1$ & $(3,3)$ & $\frac12C^{P1}_{33}$\smallskip\\
  $A2$ & $(1,1)$, $(2,2)$ & $\frac12(C^{P1}_{11}+C^{P1}_{22})+C^{P1}_{12}$\\
  $A2$ & $(1,1)$, $(3,3)$ & $\frac12(C^{P1}_{11}+C^{P1}_{33})+C^{P1}_{13}$\\
  $A2$ & $(2,2)$, $(2,3)$ & $\frac12(C^{P1}_{22}+C^{P1}_{33})+C^{P1}_{23}$\smallskip\\
  $A4$ & $(1,1)$, $(1,2)$, $(2,1)$ & $\frac12C^{P1}_{11}+2C^{P1}_{66}+2C^{P1}_{16}$\\
  $A4$ & $(1,1)$, $(1,3)$, $(3,1)$ & $\frac12C^{P1}_{11}+2C^{P1}_{55}+2C^{P1}_{15}$\\
  $A4$ & $(1,1)$, $(2,3)$, $(3,2)$ & $\frac12C^{P1}_{11}+2C^{P1}_{44}+2C^{P1}_{14}$\\
  $A4$ & $(2,2)$, $(1,2)$, $(2,1)$ & $\frac12C^{P1}_{22}+2C^{P1}_{66}+2C^{P1}_{26}$\\
  $A4$ & $(2,2)$, $(1,3)$, $(3,1)$ & $\frac12C^{P1}_{22}+2C^{P1}_{55}+2C^{P1}_{25}$\\
  $A4$ & $(2,2)$, $(2,3)$, $(3,2)$ & $\frac12C^{P1}_{22}+2C^{P1}_{44}+2C^{P1}_{24}$\\
  $A4$ & $(3,3)$, $(1,2)$, $(2,1)$ & $\frac12C^{P1}_{33}+2C^{P1}_{66}+2C^{P1}_{36}$\\
  $A4$ & $(3,3)$, $(1,3)$, $(3,1)$ & $\frac12C^{P1}_{33}+2C^{P1}_{55}+2C^{P1}_{35}$\\
  $A4$ & $(3,3)$, $(2,3)$, $(3,2)$ & $\frac12C^{P1}_{33}+2C^{P1}_{44}+2C^{P1}_{34}$
\end{tabular}
\end{ruledtabular}
\end{table}

Finally, application of the symmetry-based projection technique \cite{Moakher2006} as described in Ref.~\onlinecite{Tasnadi2012} yields the three cubic elastic constants $C_{11}$, $C_{12}$ and $C_{44}$:
\begin{gather}
  C_{11}=\frac{C^{P1}_{11}+C^{P1}_{22}+C^{P1}_{33}}3\ ,\label{eq:proj1}\\
  C_{12}=\frac{C^{P1}_{12}+C^{P1}_{13}+C^{P1}_{23}}3\ ,\\
  C_{44}=\frac{C^{P1}_{44}+C^{P1}_{55}+C^{P1}_{66}}3\ . \label{eq:proj3}
\end{gather}
The anisotropicity of the material is quantified using Zener's anisotropy ratio, $A$:
\begin{equation}
  A=\frac{2C_{44}}{C_{11}-C_{12}}\ .
\end{equation}

The stress-strain method \cite{Yu2010-vr} was used to confirm the total energy calculations of elastic constants. Additional tests were performed using 96-atom $4\times3\times4$ SQS supercell from Ref.~\onlinecite{Tasnadi2012}.

Orientation distribution function (ODF) is a convenient way to quantify the texture of polycrystals. $\text{ODF}(\alpha,\beta,\gamma)$ is a function of three Euler angles, $\alpha,\beta,\gamma$, and for particular values it gives a fraction of grains with that orientation \cite{Bunge1982-ea}. The Voigt (constant strain in all grains) and Reuss (constant stress in all grains) polycrystalline averages of elastic constants are defined as \cite{Martinschitz2009}
\begin{gather}
  C^{V}_{ijkl}=\int_{\alpha,\beta,\gamma}\text{ODF}(\alpha,\beta,\gamma)C_{ijkl}(\alpha,\beta,\gamma)\,d\alpha d\beta d\gamma\ ,\label{eq:voigt}\\
  (C^{R})^{-1}_{ijkl}=\int_{\alpha,\beta,\gamma}\text{ODF}(\alpha,\beta,\gamma)S_{ijkl}(\alpha,\beta,\gamma)\,d\alpha d\beta d\gamma\ .\label{eq:reuss}
\end{gather}
$C_{ijkl}(\alpha,\beta,\gamma)$ and $S_{ijkl}(\alpha,\beta,\gamma)$ in the above expressions are stiffness and compliance tensors, respectively, in a coordinate frame rotated by angles $\alpha$, $\beta$, and $\gamma$ with respect to the reference coordinate frame, in which both ODF and the polycrystalline elastic constants are defined. In fact, Voigt and Reuss elastic averages represents elastic response of a polycrystalline material with grain boundaries oriented parallel and perpendicular with respect to the applied stress direction. A single-valued polycrystalline elastic properties are  obtained from Hill's average
\begin{equation}
  C^H_{ijkl}=\frac12\left(C^V_{ijkl}+C^R_{ijkl}\right)\ .
\end{equation}

A commercial package LaboTex \cite{LaboSoft2006} was used to generate the ODFs describing cubic fibre texture with preferred $\langle1\,0\,0\rangle$ or $\langle1\,1\,1\rangle$ orientations, different sharpness of the distribution (quantified by the full-width at half maximum (FWHM) of the distribution\cite{Martinschitz2009}), and varying isotropic fraction.

\subsection{First principle calculations}

The quantum mechanical calculation within the framework of Density Functional Theory (DFT) were performed using Vienna Ab initio Simulation Package (VASP) \cite{Kresse1993,Kresse1996}. The exchange and correlation effects were treated using generalised gradient approximation (GGA) as parametrised by Perdew, Burke and Ernzerhof \cite{Perdew1996} and implemented in projector augmented wave pseudopotentials \cite{Blochl1994,Kresse1999}. We used plane wave cut-off of $\SI{700}{\electronvolt}$ ($\SI{500}{\electronvolt}$) and $7\times7\times7$ ($6\times6\times6$) Monkhorts-Pack $k$-point mesh for the 64 (96) atom supercells, yielding the total energy accuracy in the order of $\si{\milli\electronvolt\per\atom}$. The slightly different parameter sets are a consequence of combining results of two research groups; nevertheless, additional tests revealed that the changes in elastic constants induced by increasing the plane-wave cut-off energy from $\SI{500}{\electronvolt}$ to $\SI{700}{\electronvolt}$ and/or altering the $k$-mesh are not larger than a few GPa. All supercells were fully structurally optimised yielding energies and lattice parameters as discussed in Ref.~\onlinecite{Holec2011a}.

\section{Results}
\subsection{Single crystal elastic constants}
The single crystal elastic constants calculated using the total energy method as a function of the composition are shown in Fig.~\ref{fig:Cij}. $C_{11}$, describing the uni-axial elastic response, decreases from $\SI{522}{\giga\pascal}$ for ZrN to $ \SI{377}{\giga\pascal}$ for Zr$_{0.25}$Al$_{0.75}$N, and then increases again to $\SI{421}{\giga\pascal}$ for pure cubic AlN. On contrary, {off-diagonal shear-related  components $C_{12}$ and $C_{44}$ increase with the AlN mole fraction from $\SI{118}{\giga\pascal}$ (ZrN) to $\SI{165}{\giga\pascal}$ (AlN), and from $\SI{105}{\giga\pascal}$ (ZrN) to $\SI{306}{\giga\pascal}$ (AlN), respectively. As a results, also the Zener's anisotropy ratio, $A$, monotonically increases with the AlN mole fraction. This corresponds to a qualitative change of the directional Young's modulus distribution: the stiffest direction is $\langle1\,0\,0\rangle$ for ZrN and $\langle1\,1\,1\rangle$ for AlN \cite{Holec2012}.

The error bars show standard deviation as obtained by averaging the three elastic constants equivalent for perfect cubic material \cite{Tasnadi2012}. The relative error is below $3\%$ for $C_{11}$, around $5\%$ for $C_{12}$ and below $7\%$ for $C_{44}$ (with the exception of Zr$_{0.5}$Al$_{0.5}$N where it is $11\%$). Consequently, one can in principle rely on values obtained only for deformation e.g. along the $x$ direction to get an estimate for the $C_{11}$ with an accuracy better than $3\%$ (a value usually regarded as an acceptable deviation between theory and experiment due to various exchange-correlation effects, temperature of measurement/calculation, material quality, etc.~\cite{Frik2011-rp}).

\begin{figure}
  \includegraphics[width=\columnwidth]{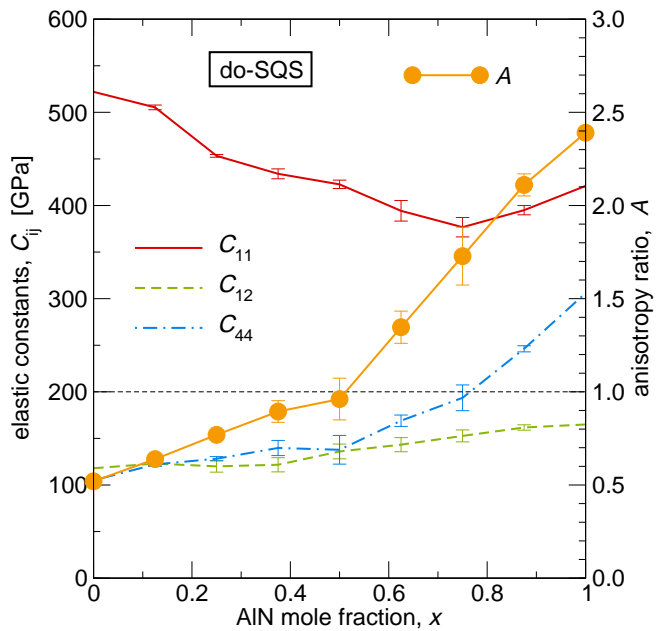}
  \caption{Single crystal cubic elastic constants $C_{11}$, $C_{12}$, and $C_{44}$, and the Zener's anisotropy ratio $A$ as functions of the AlN mole fraction in \ZrAlN, calculated using the do-SQS approach.}\label{fig:Cij}
\end{figure}

The Zener's anisotropy ratio reaches 1 for Zr$_{0.5}$Al$_{0.5}$N, implying that for this composition the alloy should have an isotropic elastic response independent of texture or crystallite orientation. This idea is further confirmed by evaluating the polycrystalline elastic constants in the next sections.

\subsection{Elastic response of isotropic polycrystal aggregates}

We begin with presenting the polycrystalline properties of isotropic aggregates, i.e. when all crystal orientations are equally probable. Figure~\ref{fig:polycrystalline} shows the compositional dependence of bulk modulus, $B$, Young's modulus, $E$, and shear modulus, $G$. The spread of the data as shown by the shaded area corresponds to the Voigt (upper) and Reuss (lower) limits. In the case of isotropic aggregates of cubic materials, i.e. when the ODF is a constant function, the Eqs.~\ref{eq:voigt} and \ref{eq:reuss} simplify to \cite{Holec2012}
\begin{gather}
  G_V=\frac{C_{11}-C_{12}+3C_{44}}5\ ,\\
  G_R=\frac{5}{4(S_{11}-S_{12})+3S_{44}}\ ,\\
  E_{\alpha}=\frac{9BG_{\alpha}}{3B+G_{\alpha}}\ ,
\end{gather}
where $\alpha=V$ or $R$ and $S_{ij}$ are elastic compliances corresponding to the elastic constants $C_{ij}$ \cite{Nye1957}.

Bulk modulus changes only a little with the composition of \ZrAlN, and the alloy is predicted to be some $10\%$ softer than the binary ZrN ($B=\SI{244}{\giga\pascal}$) and AlN ($B=\SI{250}{\giga\pascal}$). $E$ and $G$ exhibit the same behaviour, being almost constant for AlN mole fractions up to $x\approx0.6$, and only then significantly rising. The range between Voigt and Reuss limits is largest for the binary nitrides, and becomes almost zero for $x\approx0.4-0.5$. Hence for these compositions, the microstructure (lamellar or columnar) does not play a role for the resulting elastic response, as has been already stated in the previous subsection based on the Zener's anisotropy ratio.

\begin{figure}
  \includegraphics[width=\columnwidth]{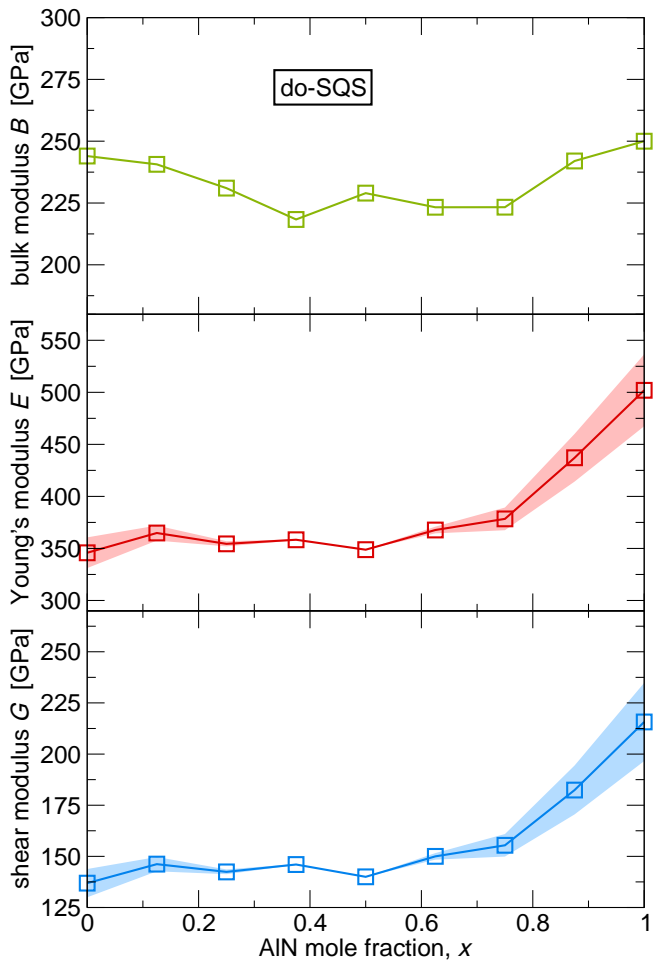}
  \caption{Bulk, Young's and shear moduli for isotropic aggregates of \ZrAlN grains as obtained using the do-SQS method.}\label{fig:polycrystalline}
\end{figure}

\begin{figure*}
  \includegraphics[width=\textwidth]{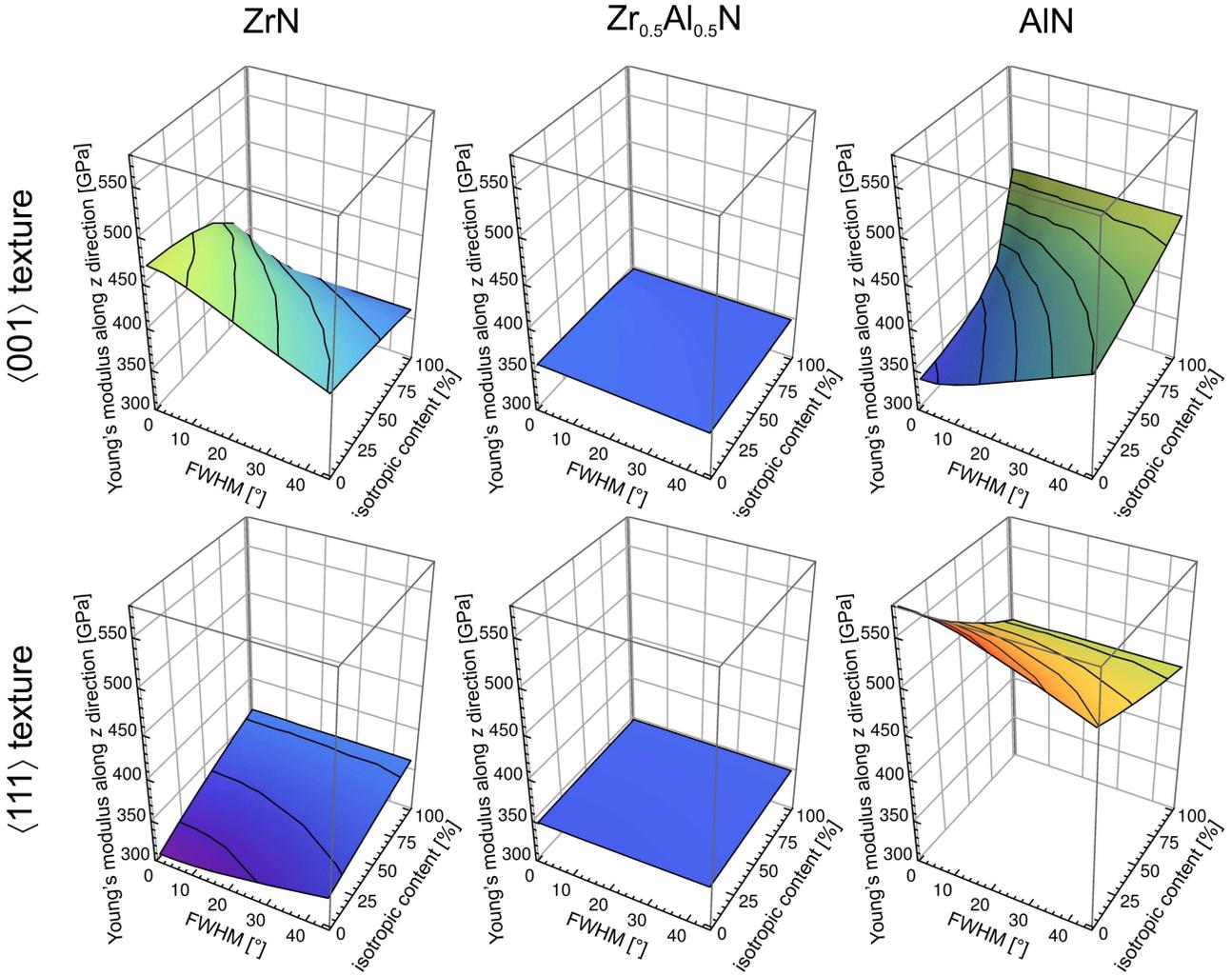}
  \caption{Dependence of the elastic response, as measured by the Young's modulus in the $z$ direction, on the texture sharpness (FWHM), content of the isotropic background, and the alloy composition. The evaluation was done for fibre textures along $\langle1\,0\,0\rangle$ and $\langle1\,1\,1\rangle$ directions.}\label{fig:texture_3D}
\end{figure*}

\subsection{Influence of fibre texture}

The elastic behaviour of real materials with always unique microstructure is, however, different from that of isotropic polycrystalline aggregates. Hard ceramic coatings typically exhibit a $\langle1\,0\,0\rangle$ or $\langle1\,1\,1\rangle$ fibre textures\cite{Mayrhofer2006}, with a fibre axis oriented perpendicular to the substrate surface, which usually develops due to the minimization of the strain energy during nucleation or as a result of the surface/interface energy minimization, respectively. Hence we have investigated the influence of a particular fibre texture containing a certain fraction of isotropic background (isotropic aggregate of grains) on the elastic Young's modulus in a direction perpendicular to the film surface ($z$ direction). Sharpness of the texture (width of the grain orientation distribution along the preferred orientation) is quantified by the full width at half maximum (FWHM) parameter \cite{Martinschitz2009}. Fig.~\ref{fig:texture_3D} shows representative results of the Hill's average of Young's modulus in the $z$ direction.

The values of the elastic constants for crystals with $\text{FWHM}=0.5^\circ$ and no isotropic background approach the single crystal directional Young's modulus in $\langle1\,0\,0\rangle$ and $\langle1\,1\,1\rangle$ directions, respectively. These values for ZrN ($E_{\langle1\,0\,0\rangle}=\SI{469}{\giga\pascal}$, $E_{\langle1\,1\,1\rangle}=\SI{307}{\giga\pascal}$) and AlN ($E_{\langle1\,0\,0\rangle}=\SI{338}{\giga\pascal}$, $E_{\langle1\,1\,1\rangle}=\SI{579}{\giga\pascal}$) exhibit the change of the softest (stiffest) direction from $\langle1\,1\,1\rangle$ ($\langle1\,0\,0\rangle$) for ZrN to $\langle1\,0\,0\rangle$ ($\langle1\,1\,1\rangle$) for AlN, as shown also earlier in Ref.~\onlinecite{Holec2012}.

Increase of the FWHM leads to decrease (increase) of the Young's modulus in the $z$-direction for ZrN with the $\langle1\,0\,0\rangle$ ($\langle1\,1\,1\rangle$) texture, as the contribution from softer (stiffer) oriented grains increases. Similar but opposite trend is obtained also for AlN. Increase of the isotropic content in the texture has qualitatively the same impact as increasing FWHM (decreasing sharpness of the texture). The Young's modulus values are independent of the FWHM parameter for $100\%$ isotropic content, since in these cases no texture fibre is present. Hence also the values in the $\langle1\,0\,0\rangle$ and $\langle1\,1\,1\rangle$ plots are the same for $100\%$ isotropic texture, and correspond to the values shown in Fig.~\ref{fig:polycrystalline}.

The shape of the Young's modulus profile changes continuously with the composition. It is worth noting that for $x=0.5$, almost flat dependence is obtained. This underlines our earlier estimates that the elastic response of a \ZrAlN solid solution around this composition is isotropic and independent of the texture (assuming the grain boundaries influence is negligible). To illustrate the compositional dependence further, we plotted in Fig.~\ref{fig:doSQS_texture} the spread between the single crystal Young's moduli in the $\langle1\,0\,0\rangle$ and $\langle1\,1\,1\rangle$ directions. It follows that the elastic response is most sensitive to the texture for the binary ZrN and AlN. Both theory \cite{Sheng2008,Holec2011a} and experiment \cite{Hasegawa2005,Rogstrom2012a,Lamni2005} suggest that the maximum AlN mole fraction in a supersaturated cubic single phase is about $x\approx0.4$. Consequently it can be stated, that addition of Al into ZrN up to its solubility limit results in the solid solution becoming steadily more elastically isotropic, hence decreasing the impact of film microstructure on the elastic behaviour.

\begin{figure}
  \includegraphics[width=\columnwidth]{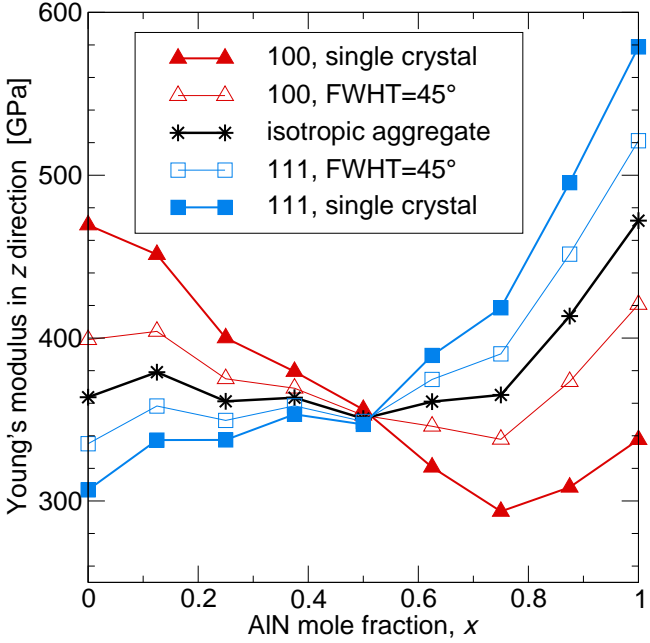}
  \caption{Compositional dependence of the Young's modulus for various textures as a function of AlN mole fraction. These data were calculated using the do-SQS cells, and suggest $x\approx0.5$ to be the texture-independent composition.}\label{fig:doSQS_texture}
\end{figure}

\section{Discussion}
\subsection{Supercell size and the method used}

A complementary calculation of the elastic constants have been performed in order to critically assess the quality of the here presented data. The results obtained with standard but larger supercells (96 atoms, $4\times3\times4$ supercell based on the fcc unit cell \cite{Tasnadi2012}) are shown together with the previously discussed elastic constants in Fig.~\ref{fig:ferenc}. It follows that there is up to $\approx13\%$ difference in the $C_{11}$ values, while the $C_{12}$ and $C_{44}$ are practically unchanged. As a consequence, the SQS-based calculation predicts Zener's anisotropy ratio $A$ to be significantly higher than the value based on the do-SQS supercell. A test calculation using do-SQS and the stress-strain method yields the same result as the total energy approach applied to the do-SQS. Another test calculation using an ordinary 64 atom SQS cell yielded $C_{11}$ $\approx2\%$ smaller than the corresponding do-SQS value. It can be therefore concluded that the discrepancies shown in Fig.~\ref{fig:ferenc} originate from the different supercell sizes (and shapes). This is similar to the behaviour of an isovalent system Ti$_{1-x}$Al$_x$N, where $C_{11}$ from 64 atom supercell is about $7.5\%$ smaller than the corresponding value calculated using a 96 atom supercell. 

Moreover, the shown 96 atom SQS data are $C_{11}^{P1}$, $C_{12}^{P1}$, and $C_{44}^{P1}$ instead of the projected elastic constants $C_{11}$, $C_{12}$, and $C_{44}$ (for which 9 elastic constants would be needed, see Eqs.~\ref{eq:proj1}--\ref{eq:proj3}). As shown in the Ref.~\onlinecite{Tasnadi2012}, when only $C_{11}^{P1}$, $C_{12}^{P1}$, and $C_{44}^{P1}$ are used to calculate $A$ (a value labelled $A_x$), $A_x$ is overestimated by $\approx18\%$ with respect to the value $A$ based on the project elastic constants for $x=0.5$. The other set of elastic constants, $C_{22}^{P1}$, $C_{23}^{P1}$, and $C_{55}^{P1}$ ($A_y$), and $C_{33}^{P1}$, $C_{13}^{P1}$, and $C_{66}^{P1}$ ($A_z$) underestimate $A$ by $9\%$ and $7\%$, respectively. On the contrary, do-SQS results in a significantly reduced the spread of the Zener's ratio. For example, for $x=0.5$ our data yield $A_x/A=1.026$, $A_x/A=0.997$, and $A_x/A=0.978$. A similar trend was observed also for the standard 64 atom SQS in Ref.~\onlinecite{Tasnadi2012}.

The supercell size together with using/not using the projection technique for $C_{ij}$ add up, and cause the Zener's anisotropy ratio, $A$, of the 96 atom supercell increase steeper (than the do-SQS value) for low AlN mole fractions, yielding $x\approx0.35$ as the composition with the nearly-isotropic response. As a consequence of the overestimated $A$ for the 96 atom SQS, the isotropic concentration is expected to be shifted to higher Al concentrations in Fig.~\ref{fig:ferenc}, hence get closer to the do-SQS predictions, when the corrected projected $C_{ij}$ values are used. 

It can be therefore concluded, that the here proposed do-SQS cells are more appropriate for a direct estimation of $C_{ij}$ (i.e., without projecting the 9 $P1$ values on the 3 cubic $C_{11}$, $C_{12}$, and $C_{44}$) than the 96 atom SQS supercell.

% This is also demonstrated by the texture-dependent spread of the Young's modulus as shown in Fig.~\ref{fig:textureFerenc} (compare with Fig.~\ref{fig:doSQS_texture}).

\begin{figure}
  \centering
  \includegraphics[width=\columnwidth]{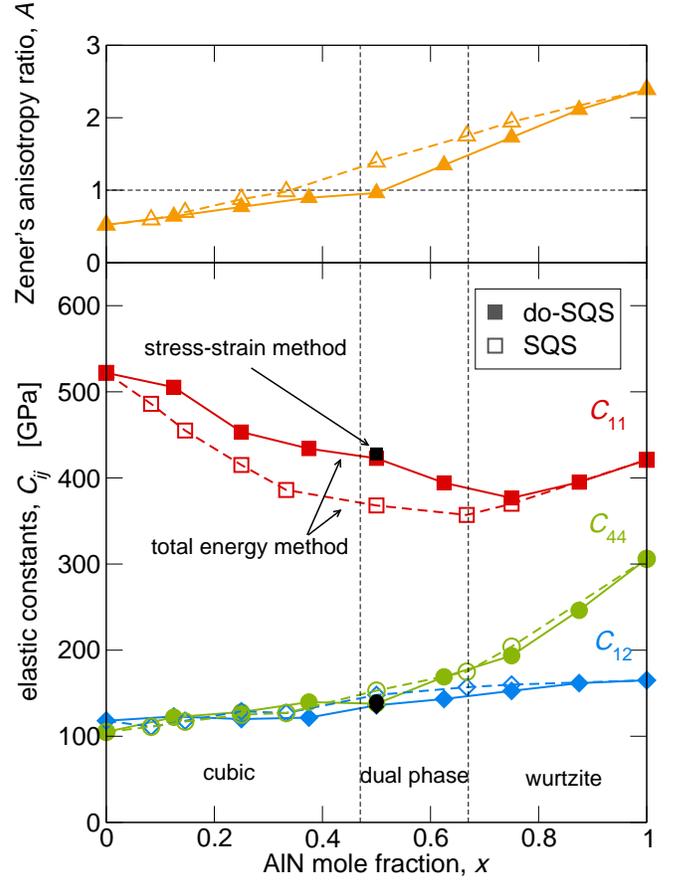}
  \caption{An overview of the supercell size and calculation method impact on the predicted single crystal elastic constants as functions of the \ZrAlN alloy composition. The elastic constants $C_{11}$, $C_{12}$, and $C_{44}$ are represented by squares, diamonds, and circles, respectively, while triangles are used for the Zener's anisotropy ratio, $A$. Full and open symbols correspond to do-SQS and SQS cells, respectively. The small black symbols represent results obtained from the stress-strain method; all other data points were calculated using the total energy method.}\label{fig:ferenc}
\end{figure}

Focusing on the single crystal elastic constants for compositions $x<0.4$ (i.e. those experimentally accessible in a single phase), the relative error from averaging $C_{11}$ from deformations along the $x$, $y$, and $z$ directions is maximum $\approx 1\%$, while it is below $\approx 6\%$ for $C_{12}$ and $C_{44}$\footnote{This is likely to be related to the way how do-SQS cells are constrained during their generation. To decrease the spread between e.g. $C_{12}$, $C_{13}$, and $C_{23}$, additional cubic symmetries have to be considered during the cell construction.}. When such accuracy is acceptable, the here proposed do-SQSs can be used to perform only one of the three symmetry equivalent deformations (i.e., $x$, $y$, or $z$ for the $C_{11}$) to obtain the respective elastic constant of the alloy.

% \begin{figure}
%   \centering
%   \includegraphics[width=\columnwidth]{ZrAlN_SQS_texture_spread.eps}
%   \caption{Compositional dependence of the Young's modulus for various textures as a function of AlN mole fraction. These data were calculated using standard SQS cells with 96 atoms. The isotropic texture-independent composition shifts to $x\approx0.35$.}\label{fig:textureFerenc}  
% \end{figure}

The off-diagonal components in rows $i=4,5,6$ and columns $j=4,5,6$ set to $\SI{0}{\giga\pascal}$ during the fitting procedure of the total energy method were confirmed to be negligibly small (below $\SI{3}{\giga\pascal}$) using the stress-strains method. Therefore, the used assumption does not influence our results.

As pointed out in the recent publications (see e.g. Refs.~\onlinecite{Caro2013,Zhou2012}), the stress-strain method proofs to be more robust than the total energy approach. It is also faster as one deformation yields several linearly independent equations for obtaining the full $6\times6$ matrix of elastic constants. It follows, that the stress-strain method performed on a standard SQS (possibly with more atoms) should be the preferred method. When a reliable inter-atomic potential exists, a molecular dynamics based approach on large SQSs \cite{VonPezold2010} may be an acceptable approach. However, in the case when neither stress tensor nor good non-DFT based method is available, we propose that our do-SQSs together with the total energy method are a suitable (CPU-time affordable) approach.

\subsection{Comparison with experimental data}

There are only a few experimental reports on the mechanical properties of \ZrAlN monolithic films in the literature. \citet{Lamni2005} reported on magnetron sputtered thin films which keep cubic structure up to AlN mole fraction $x=0.43$. The films have preferred $\langle1\,1\,1\rangle$ texture. The Young's modulus, as measured by nanoindentation, increases from $\SI{250}{\giga\pascal}$ for pure ZrN to $\SI{300}{\giga\pascal}$ for Zr$_{0.57}$Al$_{0.43}$N. The predicted Young's modulus for the preferred $\langle1\,1\,1\rangle$ texture increases from $\SI{307}{\giga\pascal}$ ($\SI{335}{\giga\pascal}$) for ZrN to $\SI{389}{\giga\pascal}$ ($\SI{375}{\giga\pascal}$) for Zr$_{0.625}$Al$_{0.375}$N for $\text{FWHM}=0.5^\circ$ ($45^\circ$) (compare with Fig.~\ref{fig:doSQS_texture}). Therefore, the trend as well as the magnitude of the increase is correctly predicted by our calculations. The somewhat softer Young's modulus experimentally is likely to be related to the presence of grain boundaries in the real microstructure as well as finite temperature during experimental measurement. The same authors also published a value of nanoindentation Young's modulus $\SI{250}{\giga\pascal}$ for the as deposited Zr$_{0.57}$Al$_{0.43}$N \cite{Sanjines2006}, which increased to $\SI{265}{\giga\pascal}$ after annealing the sample to $\SI{850}{\celsius}$. \citet{Rogstrom2012a} also observed increase in Young's modulus after annealing their arc-deposited Zr$_{0.52}$Al$_{0.48}$N to $\SI{1400}{\celsius}$, and argued that this is a consequence of increased grain size, decreased porosity and improved crystallinity of their sample. Although the grain size is not reflected in our calculations (ODF takes into account only the grain orientations, not their size nor shape), its effect, i.e. decreasing the volume fraction of the soft grain boundaries, can be intuitively foreseen. It therefore further underlines the importance of the grain boundaries (as well as the amorphous matrix present in the case of \ZrAlN) on the mechanical properties of nanocrystalline materials. This topic cannot be easily handled by the means of Density Functional Theory itself, and would require use of multi-method scale-bridging techniques (see e.g. \cite{Nikolov2010,Ma2008}).

Nevertheless, the influence of the grain boundary fraction has been experimentally proven by our results for reactively prepared Zr$_{0.65}$Al$_{0.35}$N and non-reactively prepared Zr$_{0.68}$Al$_{0.32}$N coatings possessing a single phase cubic structure and a mixed $\langle1\,1\,1\rangle$--$\langle2\,0\,0\rangle$ orientation. The reactively prepared coatings have grain sizes of $\SI{8}{\nano\meter}$ and Young's moduli of $\SI{347}{\giga\pascal}$, whereas the non-reactively prepared coatings have grain sizes of $\SI{36}{\nano\meter}$ and also larger Young's moduli of $\SI{398}{\giga\pascal}$ \cite{Paul2013}. The latter is in almost perfect agreement to our calculations for Zr$_{0.625}$Al$_{0.375}$N with $\text{FWHM}=0.5^\circ$ yielding $\SI{389}{\giga\pascal}$. 

\section{Conclusions}

The here proposed directionally-optimised SQSs seem to be a reasonable alternative to large standard SQSs for the estimation of elastic properties of alloys, in particular for systems with cubic symmetry. When well optimised, they can provide accurate single crystal elastic constants while significantly reducing the number of calculations needed by omitting the need for symmetry-based projection of $C_{ij}$.

Based on the calculated single crystal elastic constants we propose that \ZrAlN with AlN mole fraction $x\approx0.4$--$0.5$ exhibits isotropic elastic behaviour. This in particular means, that any polycrystal with this composition will have texture-independent Young's modulus. This hypothesis has been supported by explicitly evaluating the compositional dependence of Young's modulus on fibre texture orientation, its sharpness and the amount of isotropic background. The comparison with experimental data showed decent agreement with our theoretically predicted values. The small discrepancy is ascribed mainly to the influence of grain boundaries. This phenomenon, however, goes beyond the capabilities of DFT and requires a multi-scale/multi-method approach.

\begin{acknowledgments}
D.H and P.H.M. greatly acknowledge the financial support by the START Program (Y371) of the Austrian Science Fund (FWF), as well as the CPU time at the Vienna Scientific Cluster (VSC). M.F. acknowledges financial support from the Academy of Sciences of the Czech Republic through the Fellowship of Jan Evangelista Purkyn{\v e} and the access to the MetaCentrum and CERIT-SC computing facilities.
\end{acknowledgments}

\bibliographystyle{apsrev4-1}
\bibliography{Cij_paper}

\begin{thebibliography}{41}%
\makeatletter
\providecommand \@ifxundefined [1]{%
 \@ifx{#1\undefined}
}%
\providecommand \@ifnum [1]{%
 \ifnum #1\expandafter \@firstoftwo
 \else \expandafter \@secondoftwo
 \fi
}%
\providecommand \@ifx [1]{%
 \ifx #1\expandafter \@firstoftwo
 \else \expandafter \@secondoftwo
 \fi
}%
\providecommand \natexlab [1]{#1}%
\providecommand \enquote  [1]{``#1''}%
\providecommand \bibnamefont  [1]{#1}%
\providecommand \bibfnamefont [1]{#1}%
\providecommand \citenamefont [1]{#1}%
\providecommand \href@noop [0]{\@secondoftwo}%
\providecommand \href [0]{\begingroup \@sanitize@url \@href}%
\providecommand \@href[1]{\@@startlink{#1}\@@href}%
\providecommand \@@href[1]{\endgroup#1\@@endlink}%
\providecommand \@sanitize@url [0]{\catcode `\\12\catcode `\$12\catcode
  `\&12\catcode `\#12\catcode `\^12\catcode `\_12\catcode `\%12\relax}%
\providecommand \@@startlink[1]{}%
\providecommand \@@endlink[0]{}%
\providecommand \url  [0]{\begingroup\@sanitize@url \@url }%
\providecommand \@url [1]{\endgroup\@href {#1}{\urlprefix }}%
\providecommand \urlprefix  [0]{URL }%
\providecommand \Eprint [0]{\href }%
\providecommand \doibase [0]{http://dx.doi.org/}%
\providecommand \selectlanguage [0]{\@gobble}%
\providecommand \bibinfo  [0]{\@secondoftwo}%
\providecommand \bibfield  [0]{\@secondoftwo}%
\providecommand \translation [1]{[#1]}%
\providecommand \BibitemOpen [0]{}%
\providecommand \bibitemStop [0]{}%
\providecommand \bibitemNoStop [0]{.\EOS\space}%
\providecommand \EOS [0]{\spacefactor3000\relax}%
\providecommand \BibitemShut  [1]{\csname bibitem#1\endcsname}%
\let\auto@bib@innerbib\@empty
%</preamble>
\bibitem [{\citenamefont {Mayrhofer}\ \emph
  {et~al.}(2006{\natexlab{a}})\citenamefont {Mayrhofer}, \citenamefont
  {Music},\ and\ \citenamefont {Schneider}}]{Mayrhofer2006a}%
  \BibitemOpen
  \bibfield  {author} {\bibinfo {author} {\bibfnamefont {P.~H.}\ \bibnamefont
  {Mayrhofer}}, \bibinfo {author} {\bibfnamefont {D.}~\bibnamefont {Music}}, \
  and\ \bibinfo {author} {\bibfnamefont {J.~M.}\ \bibnamefont {Schneider}},\
  }\href {\doibase 10.1063/1.2360778} {\bibfield  {journal} {\bibinfo
  {journal} {Journal of Applied Physics}\ }\textbf {\bibinfo {volume} {100}},\
  \bibinfo {pages} {094906} (\bibinfo {year} {2006}{\natexlab{a}})}\BibitemShut
  {NoStop}%
\bibitem [{\citenamefont {Alling}\ \emph
  {et~al.}(2007{\natexlab{a}})\citenamefont {Alling}, \citenamefont {Marten},
  \citenamefont {Abrikosov},\ and\ \citenamefont {Karimi}}]{Alling2007}%
  \BibitemOpen
  \bibfield  {author} {\bibinfo {author} {\bibfnamefont {B.}~\bibnamefont
  {Alling}}, \bibinfo {author} {\bibfnamefont {T.}~\bibnamefont {Marten}},
  \bibinfo {author} {\bibfnamefont {I.~A.}\ \bibnamefont {Abrikosov}}, \ and\
  \bibinfo {author} {\bibfnamefont {a.}~\bibnamefont {Karimi}},\ }\href
  {\doibase 10.1063/1.2773625} {\bibfield  {journal} {\bibinfo  {journal}
  {Journal of Applied Physics}\ }\textbf {\bibinfo {volume} {102}},\ \bibinfo
  {pages} {044314} (\bibinfo {year} {2007}{\natexlab{a}})}\BibitemShut
  {NoStop}%
\bibitem [{\citenamefont {Sheng}\ \emph {et~al.}(2008)\citenamefont {Sheng},
  \citenamefont {Zhang},\ and\ \citenamefont {Veprek}}]{Sheng2008}%
  \BibitemOpen
  \bibfield  {author} {\bibinfo {author} {\bibfnamefont {S.~H.}\ \bibnamefont
  {Sheng}}, \bibinfo {author} {\bibfnamefont {R.~F.}\ \bibnamefont {Zhang}}, \
  and\ \bibinfo {author} {\bibfnamefont {S.}~\bibnamefont {Veprek}},\ }\href
  {\doibase 10.1016/j.actamat.2007.10.050} {\bibfield  {journal} {\bibinfo
  {journal} {Acta Materialia}\ }\textbf {\bibinfo {volume} {56}},\ \bibinfo
  {pages} {968} (\bibinfo {year} {2008})}\BibitemShut {NoStop}%
\bibitem [{\citenamefont {Matenoglou}\ \emph {et~al.}(2009)\citenamefont
  {Matenoglou}, \citenamefont {Lekka}, \citenamefont {Koutsokeras},
  \citenamefont {Karras}, \citenamefont {Kosmidis}, \citenamefont
  {Evangelakis},\ and\ \citenamefont {Patsalas}}]{Matenoglou2009}%
  \BibitemOpen
  \bibfield  {author} {\bibinfo {author} {\bibfnamefont {G.~M.}\ \bibnamefont
  {Matenoglou}}, \bibinfo {author} {\bibfnamefont {C.~E.}\ \bibnamefont
  {Lekka}}, \bibinfo {author} {\bibfnamefont {L.~E.}\ \bibnamefont
  {Koutsokeras}}, \bibinfo {author} {\bibfnamefont {G.}~\bibnamefont {Karras}},
  \bibinfo {author} {\bibfnamefont {C.}~\bibnamefont {Kosmidis}}, \bibinfo
  {author} {\bibfnamefont {G.~A.}\ \bibnamefont {Evangelakis}}, \ and\ \bibinfo
  {author} {\bibfnamefont {P.}~\bibnamefont {Patsalas}},\ }\href {\doibase
  10.1063/1.3131824} {\bibfield  {journal} {\bibinfo  {journal} {Journal of
  Applied Physics}\ }\textbf {\bibinfo {volume} {105}},\ \bibinfo {pages}
  {103714} (\bibinfo {year} {2009})}\BibitemShut {NoStop}%
\bibitem [{\citenamefont {Sangiovanni}\ \emph {et~al.}(2011)\citenamefont
  {Sangiovanni}, \citenamefont {Hultman},\ and\ \citenamefont
  {Chirita}}]{Sangiovanni2011}%
  \BibitemOpen
  \bibfield  {author} {\bibinfo {author} {\bibfnamefont {D.~G.}\ \bibnamefont
  {Sangiovanni}}, \bibinfo {author} {\bibfnamefont {L.}~\bibnamefont
  {Hultman}}, \ and\ \bibinfo {author} {\bibfnamefont {V.}~\bibnamefont
  {Chirita}},\ }\href {\doibase 10.1016/j.actamat.2010.12.013} {\bibfield
  {journal} {\bibinfo  {journal} {Acta Materialia}\ }\textbf {\bibinfo {volume}
  {59}},\ \bibinfo {pages} {2121} (\bibinfo {year} {2011})}\BibitemShut
  {NoStop}%
\bibitem [{\citenamefont {Holec}\ \emph {et~al.}(2013)\citenamefont {Holec},
  \citenamefont {Zhou}, \citenamefont {Rachbauer},\ and\ \citenamefont
  {Mayrhofer}}]{Holec2013a}%
  \BibitemOpen
  \bibfield  {author} {\bibinfo {author} {\bibfnamefont {D.}~\bibnamefont
  {Holec}}, \bibinfo {author} {\bibfnamefont {L.}~\bibnamefont {Zhou}},
  \bibinfo {author} {\bibfnamefont {R.}~\bibnamefont {Rachbauer}}, \ and\
  \bibinfo {author} {\bibfnamefont {P.~H.}\ \bibnamefont {Mayrhofer}},\ }\href
  {\doibase 10.1063/1.4795590} {\bibfield  {journal} {\bibinfo  {journal}
  {Journal of Applied Physics}\ }\textbf {\bibinfo {volume} {113}},\ \bibinfo
  {pages} {113510} (\bibinfo {year} {2013})}\BibitemShut {NoStop}%
\bibitem [{\citenamefont {Gonis}(2000)}]{Gonis2000-xb}%
  \BibitemOpen
  \bibfield  {author} {\bibinfo {author} {\bibfnamefont {A.}~\bibnamefont
  {Gonis}},\ }\href
  {http://books.google.at/books/about/Theoretical_Materials_Science.html?hl=&id=OL-dngEACAAJ}
  {\emph {\bibinfo {title} {Theoretical Materials Science: Tracing the
  Electronic Origins of Materials Behavior}}}\ (\bibinfo  {publisher}
  {Materials Research Society},\ \bibinfo {year} {2000})\BibitemShut {NoStop}%
\bibitem [{\citenamefont {Sanchez}(2010)}]{Sanchez2010-qb}%
  \BibitemOpen
  \bibfield  {author} {\bibinfo {author} {\bibfnamefont {J.~M.}\ \bibnamefont
  {Sanchez}},\ }\href {\doibase 10.1103/PhysRevB.81.224202} {\bibfield
  {journal} {\bibinfo  {journal} {Phys. Rev. B Condens. Matter}\ }\textbf
  {\bibinfo {volume} {81}},\ \bibinfo {pages} {224202} (\bibinfo {year}
  {2010})}\BibitemShut {NoStop}%
\bibitem [{\citenamefont {Wei}\ \emph {et~al.}(1990)\citenamefont {Wei},
  \citenamefont {Ferreira}, \citenamefont {Bernard},\ and\ \citenamefont
  {Zunger}}]{Wei1990a}%
  \BibitemOpen
  \bibfield  {author} {\bibinfo {author} {\bibfnamefont {S.-H.}\ \bibnamefont
  {Wei}}, \bibinfo {author} {\bibfnamefont {L.}~\bibnamefont {Ferreira}},
  \bibinfo {author} {\bibfnamefont {J.}~\bibnamefont {Bernard}}, \ and\
  \bibinfo {author} {\bibfnamefont {A.}~\bibnamefont {Zunger}},\ }\href
  {\doibase 10.1103/PhysRevB.42.9622} {\bibfield  {journal} {\bibinfo
  {journal} {Physical Review B}\ }\textbf {\bibinfo {volume} {42}},\ \bibinfo
  {pages} {9622} (\bibinfo {year} {1990})}\BibitemShut {NoStop}%
\bibitem [{\citenamefont {Birch}(1947)}]{Birch1947}%
  \BibitemOpen
  \bibfield  {author} {\bibinfo {author} {\bibfnamefont {F.}~\bibnamefont
  {Birch}},\ }\href {http://prola.aps.org/abstract/PR/v71/i11/p809\_1}
  {\bibfield  {journal} {\bibinfo  {journal} {Physical Review}\ }\textbf
  {\bibinfo {volume} {71}},\ \bibinfo {pages} {809} (\bibinfo {year}
  {1947})}\BibitemShut {NoStop}%
\bibitem [{\citenamefont {Holec}\ \emph {et~al.}(2012)\citenamefont {Holec},
  \citenamefont {Fri\'{a}k}, \citenamefont {Neugebauer},\ and\ \citenamefont
  {Mayrhofer}}]{Holec2012}%
  \BibitemOpen
  \bibfield  {author} {\bibinfo {author} {\bibfnamefont {D.}~\bibnamefont
  {Holec}}, \bibinfo {author} {\bibfnamefont {M.}~\bibnamefont {Fri\'{a}k}},
  \bibinfo {author} {\bibfnamefont {J.}~\bibnamefont {Neugebauer}}, \ and\
  \bibinfo {author} {\bibfnamefont {P.~H.}\ \bibnamefont {Mayrhofer}},\ }\href
  {\doibase 10.1103/PhysRevB.85.064101} {\bibfield  {journal} {\bibinfo
  {journal} {Physical Review B}\ }\textbf {\bibinfo {volume} {85}},\ \bibinfo
  {pages} {064101} (\bibinfo {year} {2012})}\BibitemShut {NoStop}%
\bibitem [{\citenamefont {Caro}\ \emph {et~al.}(2013)\citenamefont {Caro},
  \citenamefont {Schulz},\ and\ \citenamefont {O'Reilly}}]{Caro2013}%
  \BibitemOpen
  \bibfield  {author} {\bibinfo {author} {\bibfnamefont {M.~A.}\ \bibnamefont
  {Caro}}, \bibinfo {author} {\bibfnamefont {S.}~\bibnamefont {Schulz}}, \ and\
  \bibinfo {author} {\bibfnamefont {E.~P.}\ \bibnamefont {O'Reilly}},\ }\href
  {\doibase 10.1088/0953-8984/25/2/025803} {\bibfield  {journal} {\bibinfo
  {journal} {Journal of physics. Condensed matter : an Institute of Physics
  journal}\ }\textbf {\bibinfo {volume} {25}},\ \bibinfo {pages} {025803}
  (\bibinfo {year} {2013})}\BibitemShut {NoStop}%
\bibitem [{\citenamefont {von Pezold}\ \emph {et~al.}(2010)\citenamefont {von
  Pezold}, \citenamefont {Dick}, \citenamefont {Fri\'{a}k},\ and\ \citenamefont
  {Neugebauer}}]{VonPezold2010}%
  \BibitemOpen
  \bibfield  {author} {\bibinfo {author} {\bibfnamefont {J.}~\bibnamefont {von
  Pezold}}, \bibinfo {author} {\bibfnamefont {A.}~\bibnamefont {Dick}},
  \bibinfo {author} {\bibfnamefont {M.}~\bibnamefont {Fri\'{a}k}}, \ and\
  \bibinfo {author} {\bibfnamefont {J.}~\bibnamefont {Neugebauer}},\ }\href
  {\doibase 10.1103/PhysRevB.81.094203} {\bibfield  {journal} {\bibinfo
  {journal} {Physical Review B}\ }\textbf {\bibinfo {volume} {81}},\ \bibinfo
  {pages} {094203} (\bibinfo {year} {2010})}\BibitemShut {NoStop}%
\bibitem [{\citenamefont {Moakher}\ and\ \citenamefont
  {Norris}(2006)}]{Moakher2006}%
  \BibitemOpen
  \bibfield  {author} {\bibinfo {author} {\bibfnamefont {M.}~\bibnamefont
  {Moakher}}\ and\ \bibinfo {author} {\bibfnamefont {A.~N.}\ \bibnamefont
  {Norris}},\ }\href {\doibase 10.1007/s10659-006-9082-0} {\bibfield  {journal}
  {\bibinfo  {journal} {Journal of Elasticity}\ }\textbf {\bibinfo {volume}
  {85}},\ \bibinfo {pages} {215} (\bibinfo {year} {2006})}\BibitemShut
  {NoStop}%
\bibitem [{\citenamefont {Tasn\'{a}di}\ \emph {et~al.}(2010)\citenamefont
  {Tasn\'{a}di}, \citenamefont {Abrikosov}, \citenamefont {Rogstr\"{o}m},
  \citenamefont {Almer}, \citenamefont {Johansson},\ and\ \citenamefont
  {Od\'{e}n}}]{Tasnadi2010a}%
  \BibitemOpen
  \bibfield  {author} {\bibinfo {author} {\bibfnamefont {F.}~\bibnamefont
  {Tasn\'{a}di}}, \bibinfo {author} {\bibfnamefont {I.~A.}\ \bibnamefont
  {Abrikosov}}, \bibinfo {author} {\bibfnamefont {L.}~\bibnamefont
  {Rogstr\"{o}m}}, \bibinfo {author} {\bibfnamefont {J.}~\bibnamefont {Almer}},
  \bibinfo {author} {\bibfnamefont {M.~P.}\ \bibnamefont {Johansson}}, \ and\
  \bibinfo {author} {\bibfnamefont {M.}~\bibnamefont {Od\'{e}n}},\ }\href
  {\doibase 10.1063/1.3524502} {\bibfield  {journal} {\bibinfo  {journal}
  {Applied Physics Letters}\ }\textbf {\bibinfo {volume} {97}},\ \bibinfo
  {pages} {231902} (\bibinfo {year} {2010})}\BibitemShut {NoStop}%
\bibitem [{\citenamefont {Tasn\'{a}di}\ \emph {et~al.}(2012)\citenamefont
  {Tasn\'{a}di}, \citenamefont {Od\'{e}n},\ and\ \citenamefont
  {Abrikosov}}]{Tasnadi2012}%
  \BibitemOpen
  \bibfield  {author} {\bibinfo {author} {\bibfnamefont {F.}~\bibnamefont
  {Tasn\'{a}di}}, \bibinfo {author} {\bibfnamefont {M.}~\bibnamefont
  {Od\'{e}n}}, \ and\ \bibinfo {author} {\bibfnamefont {I.~A.}\ \bibnamefont
  {Abrikosov}},\ }\href {\doibase 10.1103/PhysRevB.85.144112} {\bibfield
  {journal} {\bibinfo  {journal} {Physical Review B}\ }\textbf {\bibinfo
  {volume} {85}},\ \bibinfo {pages} {1} (\bibinfo {year} {2012})}\BibitemShut
  {NoStop}%
\bibitem [{\citenamefont {Chen}\ \emph {et~al.}(2011)\citenamefont {Chen},
  \citenamefont {Holec}, \citenamefont {Du},\ and\ \citenamefont
  {Mayrhofer}}]{Chen2011}%
  \BibitemOpen
  \bibfield  {author} {\bibinfo {author} {\bibfnamefont {L.}~\bibnamefont
  {Chen}}, \bibinfo {author} {\bibfnamefont {D.}~\bibnamefont {Holec}},
  \bibinfo {author} {\bibfnamefont {Y.}~\bibnamefont {Du}}, \ and\ \bibinfo
  {author} {\bibfnamefont {P.~H.}\ \bibnamefont {Mayrhofer}},\ }\href {\doibase
  10.1016/j.tsf.2011.03.139} {\bibfield  {journal} {\bibinfo  {journal} {Thin
  Solid Films}\ }\textbf {\bibinfo {volume} {519}},\ \bibinfo {pages} {5503}
  (\bibinfo {year} {2011})}\BibitemShut {NoStop}%
\bibitem [{\citenamefont {M\"{u}ller}(2003)}]{Muller2003}%
  \BibitemOpen
  \bibfield  {author} {\bibinfo {author} {\bibfnamefont {S.}~\bibnamefont
  {M\"{u}ller}},\ }\href {\doibase 10.1088/0953-8984/15/34/201} {\bibfield
  {journal} {\bibinfo  {journal} {Journal of Physics: Condensed Matter}\
  }\textbf {\bibinfo {volume} {15}},\ \bibinfo {pages} {R1429} (\bibinfo {year}
  {2003})}\BibitemShut {NoStop}%
\bibitem [{\citenamefont {Alling}\ \emph
  {et~al.}(2007{\natexlab{b}})\citenamefont {Alling}, \citenamefont {Ruban},
  \citenamefont {Karimi}, \citenamefont {Peil}, \citenamefont {Simak},
  \citenamefont {Hultman},\ and\ \citenamefont {Abrikosov}}]{Alling2007a}%
  \BibitemOpen
  \bibfield  {author} {\bibinfo {author} {\bibfnamefont {B.}~\bibnamefont
  {Alling}}, \bibinfo {author} {\bibfnamefont {A.~V.}\ \bibnamefont {Ruban}},
  \bibinfo {author} {\bibfnamefont {A.}~\bibnamefont {Karimi}}, \bibinfo
  {author} {\bibfnamefont {O.}~\bibnamefont {Peil}}, \bibinfo {author}
  {\bibfnamefont {S.}~\bibnamefont {Simak}}, \bibinfo {author} {\bibfnamefont
  {L.}~\bibnamefont {Hultman}}, \ and\ \bibinfo {author} {\bibfnamefont
  {I.~A.}\ \bibnamefont {Abrikosov}},\ }\href {\doibase
  10.1103/PhysRevB.75.045123} {\bibfield  {journal} {\bibinfo  {journal}
  {Physical Review B}\ }\textbf {\bibinfo {volume} {75}},\ \bibinfo {pages}
  {045123} (\bibinfo {year} {2007}{\natexlab{b}})}\BibitemShut {NoStop}%
\bibitem [{\citenamefont {Yu}\ \emph {et~al.}(2010)\citenamefont {Yu},
  \citenamefont {Zhu},\ and\ \citenamefont {Ye}}]{Yu2010-vr}%
  \BibitemOpen
  \bibfield  {author} {\bibinfo {author} {\bibfnamefont {R.}~\bibnamefont
  {Yu}}, \bibinfo {author} {\bibfnamefont {J.}~\bibnamefont {Zhu}}, \ and\
  \bibinfo {author} {\bibfnamefont {H.}~\bibnamefont {Ye}},\ }\href {\doibase
  10.1016/j.cpc.2009.11.017} {\bibfield  {journal} {\bibinfo  {journal}
  {Comput. Phys. Commun.}\ }\textbf {\bibinfo {volume} {181}},\ \bibinfo
  {pages} {671} (\bibinfo {year} {2010})}\BibitemShut {NoStop}%
\bibitem [{\citenamefont {Bunge}(1982)}]{Bunge1982-ea}%
  \BibitemOpen
  \bibfield  {author} {\bibinfo {author} {\bibfnamefont {H.~J.}\ \bibnamefont
  {Bunge}},\ }\href {http://books.google.at/books?id=ODdRAAAAMAAJ} {\emph
  {\bibinfo {title} {Texture analysis in materials science: mathematical
  methods}}}\ (\bibinfo  {publisher} {Butterworths},\ \bibinfo {year}
  {1982})\BibitemShut {NoStop}%
\bibitem [{\citenamefont {Martinschitz}\ \emph {et~al.}(2009)\citenamefont
  {Martinschitz}, \citenamefont {Daniel}, \citenamefont {Mitterer},\ and\
  \citenamefont {Keckes}}]{Martinschitz2009}%
  \BibitemOpen
  \bibfield  {author} {\bibinfo {author} {\bibfnamefont {K.~J.}\ \bibnamefont
  {Martinschitz}}, \bibinfo {author} {\bibfnamefont {R.}~\bibnamefont
  {Daniel}}, \bibinfo {author} {\bibfnamefont {C.}~\bibnamefont {Mitterer}}, \
  and\ \bibinfo {author} {\bibfnamefont {J.}~\bibnamefont {Keckes}},\ }\href
  {\doibase 10.1107/S0021889809011807} {\bibfield  {journal} {\bibinfo
  {journal} {Journal of applied crystallography}\ }\textbf {\bibinfo {volume}
  {42}},\ \bibinfo {pages} {416} (\bibinfo {year} {2009})}\BibitemShut
  {NoStop}%
\bibitem [{\citenamefont {LaboSoft}(2006)}]{LaboSoft2006}%
  \BibitemOpen
  \bibfield  {author} {\bibinfo {author} {\bibnamefont {LaboSoft}},\ }\href
  {http://www.labosoft.com.pl/} {\enquote {\bibinfo {title} {{LaboTex}},}\ }
  (\bibinfo {year} {2006})\BibitemShut {NoStop}%
\bibitem [{\citenamefont {Kresse}\ and\ \citenamefont
  {Hafner}(1993)}]{Kresse1993}%
  \BibitemOpen
  \bibfield  {author} {\bibinfo {author} {\bibfnamefont {G.}~\bibnamefont
  {Kresse}}\ and\ \bibinfo {author} {\bibfnamefont {J.}~\bibnamefont
  {Hafner}},\ }\href {\doibase 10.1103/PhysRevB.47.558} {\bibfield  {journal}
  {\bibinfo  {journal} {Physical Review B}\ }\textbf {\bibinfo {volume} {47}},\
  \bibinfo {pages} {558} (\bibinfo {year} {1993})}\BibitemShut {NoStop}%
\bibitem [{\citenamefont {Kresse}\ and\ \citenamefont
  {Furthm\"{u}ller}(1996)}]{Kresse1996}%
  \BibitemOpen
  \bibfield  {author} {\bibinfo {author} {\bibfnamefont {G.}~\bibnamefont
  {Kresse}}\ and\ \bibinfo {author} {\bibfnamefont {J.}~\bibnamefont
  {Furthm\"{u}ller}},\ }\href {\doibase 10.1103/PhysRevB.54.11169} {\bibfield
  {journal} {\bibinfo  {journal} {Physical Review B}\ }\textbf {\bibinfo
  {volume} {54}},\ \bibinfo {pages} {11169} (\bibinfo {year}
  {1996})}\BibitemShut {NoStop}%
\bibitem [{\citenamefont {Perdew}\ \emph {et~al.}(1996)\citenamefont {Perdew},
  \citenamefont {Burke},\ and\ \citenamefont {Ernzerhof}}]{Perdew1996}%
  \BibitemOpen
  \bibfield  {author} {\bibinfo {author} {\bibfnamefont {J.~P.}\ \bibnamefont
  {Perdew}}, \bibinfo {author} {\bibfnamefont {K.}~\bibnamefont {Burke}}, \
  and\ \bibinfo {author} {\bibfnamefont {M.}~\bibnamefont {Ernzerhof}},\ }\href
  {\doibase 10.1103/PhysRevLett.77.3865} {\bibfield  {journal} {\bibinfo
  {journal} {Physical Review Letters}\ }\textbf {\bibinfo {volume} {77}},\
  \bibinfo {pages} {3865} (\bibinfo {year} {1996})}\BibitemShut {NoStop}%
\bibitem [{\citenamefont {Bl\"{o}chl}(1994)}]{Blochl1994}%
  \BibitemOpen
  \bibfield  {author} {\bibinfo {author} {\bibfnamefont {P.~E.}\ \bibnamefont
  {Bl\"{o}chl}},\ }\href {\doibase 10.1103/PhysRevB.50.17953} {\bibfield
  {journal} {\bibinfo  {journal} {Physical Review B}\ }\textbf {\bibinfo
  {volume} {50}},\ \bibinfo {pages} {17953} (\bibinfo {year}
  {1994})}\BibitemShut {NoStop}%
\bibitem [{\citenamefont {Kresse}\ and\ \citenamefont
  {Joubert}(1999)}]{Kresse1999}%
  \BibitemOpen
  \bibfield  {author} {\bibinfo {author} {\bibfnamefont {G.}~\bibnamefont
  {Kresse}}\ and\ \bibinfo {author} {\bibfnamefont {D.}~\bibnamefont
  {Joubert}},\ }\href {\doibase 10.1103/PhysRevB.59.1758} {\bibfield  {journal}
  {\bibinfo  {journal} {Physical Review B}\ }\textbf {\bibinfo {volume} {59}},\
  \bibinfo {pages} {1758} (\bibinfo {year} {1999})}\BibitemShut {NoStop}%
\bibitem [{\citenamefont {Holec}\ \emph {et~al.}(2011)\citenamefont {Holec},
  \citenamefont {Rachbauer}, \citenamefont {Chen}, \citenamefont {Wang},
  \citenamefont {Luef},\ and\ \citenamefont {Mayrhofer}}]{Holec2011a}%
  \BibitemOpen
  \bibfield  {author} {\bibinfo {author} {\bibfnamefont {D.}~\bibnamefont
  {Holec}}, \bibinfo {author} {\bibfnamefont {R.}~\bibnamefont {Rachbauer}},
  \bibinfo {author} {\bibfnamefont {L.}~\bibnamefont {Chen}}, \bibinfo {author}
  {\bibfnamefont {L.}~\bibnamefont {Wang}}, \bibinfo {author} {\bibfnamefont
  {D.}~\bibnamefont {Luef}}, \ and\ \bibinfo {author} {\bibfnamefont
  {P.~P.~H.}\ \bibnamefont {Mayrhofer}},\ }\href {\doibase
  10.1016/j.surfcoat.2011.09.019} {\bibfield  {journal} {\bibinfo  {journal}
  {Surface and Coatings Technology}\ }\textbf {\bibinfo {volume} {206}},\
  \bibinfo {pages} {1698} (\bibinfo {year} {2011})}\BibitemShut {NoStop}%
\bibitem [{\citenamefont {Fri\'{a}k}\ \emph {et~al.}(2011)\citenamefont
  {Fri\'{a}k}, \citenamefont {Hickel}, \citenamefont {Grabowski}, \citenamefont
  {Lymperakis}, \citenamefont {Udyansky}, \citenamefont {Dick}, \citenamefont
  {Ma}, \citenamefont {Roters}, \citenamefont {Zhu}, \citenamefont {Schlieter},
  \citenamefont {K{\"{u}}hn}, \citenamefont {Ebrahimi}, \citenamefont
  {Lebensohn}, \citenamefont {Holec}, \citenamefont {Eckert}, \citenamefont
  {Emmerich}, \citenamefont {Raabe},\ and\ \citenamefont
  {Neugebauer}}]{Frik2011-rp}%
  \BibitemOpen
  \bibfield  {author} {\bibinfo {author} {\bibfnamefont {M.}~\bibnamefont
  {Fri\'{a}k}}, \bibinfo {author} {\bibfnamefont {T.}~\bibnamefont {Hickel}},
  \bibinfo {author} {\bibfnamefont {B.}~\bibnamefont {Grabowski}}, \bibinfo
  {author} {\bibfnamefont {L.}~\bibnamefont {Lymperakis}}, \bibinfo {author}
  {\bibfnamefont {A.}~\bibnamefont {Udyansky}}, \bibinfo {author}
  {\bibfnamefont {A.}~\bibnamefont {Dick}}, \bibinfo {author} {\bibfnamefont
  {D.}~\bibnamefont {Ma}}, \bibinfo {author} {\bibfnamefont {F.}~\bibnamefont
  {Roters}}, \bibinfo {author} {\bibfnamefont {L.-F.}\ \bibnamefont {Zhu}},
  \bibinfo {author} {\bibfnamefont {A.}~\bibnamefont {Schlieter}}, \bibinfo
  {author} {\bibfnamefont {U.}~\bibnamefont {K{\"{u}}hn}}, \bibinfo {author}
  {\bibfnamefont {Z.}~\bibnamefont {Ebrahimi}}, \bibinfo {author}
  {\bibfnamefont {R.~A.}\ \bibnamefont {Lebensohn}}, \bibinfo {author}
  {\bibfnamefont {D.}~\bibnamefont {Holec}}, \bibinfo {author} {\bibfnamefont
  {J.}~\bibnamefont {Eckert}}, \bibinfo {author} {\bibfnamefont
  {H.}~\bibnamefont {Emmerich}}, \bibinfo {author} {\bibfnamefont
  {D.}~\bibnamefont {Raabe}}, \ and\ \bibinfo {author} {\bibfnamefont
  {J.}~\bibnamefont {Neugebauer}},\ }\href {\doibase
  10.1140/epjp/i2011-11101-2} {\bibfield  {journal} {\bibinfo  {journal} {The
  European Physical Journal Plus}\ }\textbf {\bibinfo {volume} {126}},\
  \bibinfo {pages} {1} (\bibinfo {year} {2011})}\BibitemShut {NoStop}%
\bibitem [{\citenamefont {Nye}(1957)}]{Nye1957}%
  \BibitemOpen
  \bibfield  {author} {\bibinfo {author} {\bibfnamefont {J.}~\bibnamefont
  {Nye}},\ }\href
  {http://scholar.google.com/scholar?hl=en\&btnG=Search\&q=intitle:physical+properties+of+crystals\#0}
  {\emph {\bibinfo {title} {{Physical properties of crystals}}}},\ \bibinfo
  {edition} {1st}\ ed.\ (\bibinfo  {publisher} {Clarendon Press},\ \bibinfo
  {address} {Oxford},\ \bibinfo {year} {1957})\BibitemShut {NoStop}%
\bibitem [{\citenamefont {Mayrhofer}\ \emph
  {et~al.}(2006{\natexlab{b}})\citenamefont {Mayrhofer}, \citenamefont
  {Mitterer}, \citenamefont {Hultman},\ and\ \citenamefont
  {Clemens}}]{Mayrhofer2006}%
  \BibitemOpen
  \bibfield  {author} {\bibinfo {author} {\bibfnamefont {P.~H.}\ \bibnamefont
  {Mayrhofer}}, \bibinfo {author} {\bibfnamefont {C.}~\bibnamefont {Mitterer}},
  \bibinfo {author} {\bibfnamefont {L.}~\bibnamefont {Hultman}}, \ and\
  \bibinfo {author} {\bibfnamefont {H.}~\bibnamefont {Clemens}},\ }\href
  {\doibase 10.1016/j.pmatsci.2006.02.002} {\bibfield  {journal} {\bibinfo
  {journal} {Progress in Materials Science}\ }\textbf {\bibinfo {volume}
  {51}},\ \bibinfo {pages} {1032} (\bibinfo {year}
  {2006}{\natexlab{b}})}\BibitemShut {NoStop}%
\bibitem [{\citenamefont {Hasegawa}\ \emph {et~al.}(2005)\citenamefont
  {Hasegawa}, \citenamefont {Kawate},\ and\ \citenamefont
  {Suzuki}}]{Hasegawa2005}%
  \BibitemOpen
  \bibfield  {author} {\bibinfo {author} {\bibfnamefont {H.}~\bibnamefont
  {Hasegawa}}, \bibinfo {author} {\bibfnamefont {M.}~\bibnamefont {Kawate}}, \
  and\ \bibinfo {author} {\bibfnamefont {T.}~\bibnamefont {Suzuki}},\ }\href
  {\doibase 10.1016/j.surfcoat.2004.08.208} {\bibfield  {journal} {\bibinfo
  {journal} {Surface and Coatings Technology}\ }\textbf {\bibinfo {volume}
  {200}},\ \bibinfo {pages} {2409} (\bibinfo {year} {2005})}\BibitemShut
  {NoStop}%
\bibitem [{\citenamefont {Rogstr\"{o}m}\ \emph {et~al.}(2012)\citenamefont
  {Rogstr\"{o}m}, \citenamefont {Ahlgren}, \citenamefont {Almer}, \citenamefont
  {Hultman},\ and\ \citenamefont {Od\'{e}n}}]{Rogstrom2012a}%
  \BibitemOpen
  \bibfield  {author} {\bibinfo {author} {\bibfnamefont {L.}~\bibnamefont
  {Rogstr\"{o}m}}, \bibinfo {author} {\bibfnamefont {M.}~\bibnamefont
  {Ahlgren}}, \bibinfo {author} {\bibfnamefont {J.}~\bibnamefont {Almer}},
  \bibinfo {author} {\bibfnamefont {L.}~\bibnamefont {Hultman}}, \ and\
  \bibinfo {author} {\bibfnamefont {M.}~\bibnamefont {Od\'{e}n}},\ }\href
  {\doibase 10.1557/jmr.2012.122} {\bibfield  {journal} {\bibinfo  {journal}
  {Journal of Materials Research}\ }\textbf {\bibinfo {volume} {27}},\ \bibinfo
  {pages} {1716} (\bibinfo {year} {2012})}\BibitemShut {NoStop}%
\bibitem [{\citenamefont {Lamni}\ \emph {et~al.}(2005)\citenamefont {Lamni},
  \citenamefont {Sanjinés}, \citenamefont {Parlinska-Wojtan}, \citenamefont
  {Karimi},\ and\ \citenamefont {Lévy}}]{Lamni2005}%
  \BibitemOpen
  \bibfield  {author} {\bibinfo {author} {\bibfnamefont {R.}~\bibnamefont
  {Lamni}}, \bibinfo {author} {\bibfnamefont {R.}~\bibnamefont {Sanjinés}},
  \bibinfo {author} {\bibfnamefont {M.}~\bibnamefont {Parlinska-Wojtan}},
  \bibinfo {author} {\bibfnamefont {A.}~\bibnamefont {Karimi}}, \ and\ \bibinfo
  {author} {\bibfnamefont {F.}~\bibnamefont {Lévy}},\ }\href {\doibase
  10.1116/1.1924579} {\bibfield  {journal} {\bibinfo  {journal} {Journal of
  Vacuum Science \& Technology A: Vacuum, Surfaces, and Films}\ }\textbf
  {\bibinfo {volume} {23}},\ \bibinfo {pages} {593} (\bibinfo {year}
  {2005})}\BibitemShut {NoStop}%
\bibitem [{Note1()}]{Note1}%
  \BibitemOpen
  \bibinfo {note} {This is likely to be related to the way how do-SQS cells are
  constrained during their generation. To decrease the spread between e.g.
  $C_{12}$, $C_{13}$, and $C_{23}$, additional cubic symmetries have to be
  considered during the cell construction.}\BibitemShut {Stop}%
\bibitem [{\citenamefont {Zhou}\ \emph {et~al.}(2013)\citenamefont {Zhou},
  \citenamefont {Holec},\ and\ \citenamefont {Mayrhofer}}]{Zhou2012}%
  \BibitemOpen
  \bibfield  {author} {\bibinfo {author} {\bibfnamefont {L.}~\bibnamefont
  {Zhou}}, \bibinfo {author} {\bibfnamefont {D.}~\bibnamefont {Holec}}, \ and\
  \bibinfo {author} {\bibfnamefont {P.~H.}\ \bibnamefont {Mayrhofer}},\ }\href
  {\doibase 10.1063/1.4789378} {\bibfield  {journal} {\bibinfo  {journal}
  {Journal of Applied Physics}\ }\textbf {\bibinfo {volume} {113}},\ \bibinfo
  {pages} {043511} (\bibinfo {year} {2013})}\BibitemShut {NoStop}%
\bibitem [{\citenamefont {Sanjin\'{e}s}\ \emph {et~al.}(2006)\citenamefont
  {Sanjin\'{e}s}, \citenamefont {Sandu}, \citenamefont {Lamni},\ and\
  \citenamefont {L\'{e}vy}}]{Sanjines2006}%
  \BibitemOpen
  \bibfield  {author} {\bibinfo {author} {\bibfnamefont {R.}~\bibnamefont
  {Sanjin\'{e}s}}, \bibinfo {author} {\bibfnamefont {C.~S.}\ \bibnamefont
  {Sandu}}, \bibinfo {author} {\bibfnamefont {R.}~\bibnamefont {Lamni}}, \ and\
  \bibinfo {author} {\bibfnamefont {F.}~\bibnamefont {L\'{e}vy}},\ }\href
  {\doibase 10.1016/j.surfcoat.2005.11.113} {\bibfield  {journal} {\bibinfo
  {journal} {Surface and Coatings Technology}\ }\textbf {\bibinfo {volume}
  {200}},\ \bibinfo {pages} {6308} (\bibinfo {year} {2006})}\BibitemShut
  {NoStop}%
\bibitem [{\citenamefont {Nikolov}\ \emph {et~al.}(2010)\citenamefont
  {Nikolov}, \citenamefont {Petrov}, \citenamefont {Lymperakis}, \citenamefont
  {Fri\'{a}k}, \citenamefont {Sachs}, \citenamefont {Fabritius}, \citenamefont
  {Raabe},\ and\ \citenamefont {Neugebauer}}]{Nikolov2010}%
  \BibitemOpen
  \bibfield  {author} {\bibinfo {author} {\bibfnamefont {S.}~\bibnamefont
  {Nikolov}}, \bibinfo {author} {\bibfnamefont {M.}~\bibnamefont {Petrov}},
  \bibinfo {author} {\bibfnamefont {L.}~\bibnamefont {Lymperakis}}, \bibinfo
  {author} {\bibfnamefont {M.}~\bibnamefont {Fri\'{a}k}}, \bibinfo {author}
  {\bibfnamefont {C.}~\bibnamefont {Sachs}}, \bibinfo {author} {\bibfnamefont
  {H.-O.}\ \bibnamefont {Fabritius}}, \bibinfo {author} {\bibfnamefont
  {D.}~\bibnamefont {Raabe}}, \ and\ \bibinfo {author} {\bibfnamefont
  {J.}~\bibnamefont {Neugebauer}},\ }\href {\doibase 10.1002/adma.200902019}
  {\bibfield  {journal} {\bibinfo  {journal} {Advanced materials (Deerfield
  Beach, Fla.)}\ }\textbf {\bibinfo {volume} {22}},\ \bibinfo {pages} {519}
  (\bibinfo {year} {2010})}\BibitemShut {NoStop}%
\bibitem [{\citenamefont {Ma}\ \emph {et~al.}(2008)\citenamefont {Ma},
  \citenamefont {Fri\'{a}k}, \citenamefont {Neugebauer}, \citenamefont
  {Raabe},\ and\ \citenamefont {Roters}}]{Ma2008}%
  \BibitemOpen
  \bibfield  {author} {\bibinfo {author} {\bibfnamefont {D.}~\bibnamefont
  {Ma}}, \bibinfo {author} {\bibfnamefont {M.}~\bibnamefont {Fri\'{a}k}},
  \bibinfo {author} {\bibfnamefont {J.}~\bibnamefont {Neugebauer}}, \bibinfo
  {author} {\bibfnamefont {D.}~\bibnamefont {Raabe}}, \ and\ \bibinfo {author}
  {\bibfnamefont {F.}~\bibnamefont {Roters}},\ }\href {\doibase
  10.1002/pssb.200844227} {\bibfield  {journal} {\bibinfo  {journal} {Physica
  Status Solidi (B)}\ }\textbf {\bibinfo {volume} {245}},\ \bibinfo {pages}
  {2642} (\bibinfo {year} {2008})}\BibitemShut {NoStop}%
\bibitem [{\citenamefont {Mayrhofer}\ \emph {et~al.}(2014)\citenamefont
  {Mayrhofer}, \citenamefont {Sonnleitner}, \citenamefont {Bartosik},\ and\
  \citenamefont {Holec}}]{Paul2013}%
  \BibitemOpen
  \bibfield  {author} {\bibinfo {author} {\bibfnamefont {P.}~\bibnamefont
  {Mayrhofer}}, \bibinfo {author} {\bibfnamefont {D.}~\bibnamefont
  {Sonnleitner}}, \bibinfo {author} {\bibfnamefont {M.}~\bibnamefont
  {Bartosik}}, \ and\ \bibinfo {author} {\bibfnamefont {D.}~\bibnamefont
  {Holec}},\ }\href {\doibase 10.1016/j.surfcoat.2014.01.049} {\bibfield
  {journal} {\bibinfo  {journal} {Surf. Coat. Technol.}\ }\textbf {\bibinfo
  {volume} {244}},\ \bibinfo {pages} {52} (\bibinfo {year} {2014})}\BibitemShut
  {NoStop}%
\end{thebibliography}%

\appendix
\section{Description of supercells}

\begin{table}
  \caption{Arrangement of atoms on the metallic sublattice in the supercells. Additional 32 position according to the B1 structure are occupied by N atoms. Compositions with $x_A>0.5$ are obtained by interchanging $A$ and $B$ atoms in a $(1-x_A)$ supercell.}
  \label{tab:supercells}
  \begin{ruledtabular}
  \begin{tabular}{ccc cccc}
    \multicolumn{3}{c}{site coordinates} & \multicolumn{4}{c}{mole fraction, $x_A$}\\
    $x$	& $y$	& $z$	& $0.125$	& $0.25$	& $0.375$	& $0.5$\\\hline
    0	& 0	& 0	& B	& B	& B	& A\\
    0	& 0	& 0.5	& B	& B	& B	& A\\
    0	& 0.25	& 0.25	& A	& B	& B	& B\\
    0	& 0.25	& 0.75	& B	& B	& B	& A\\
    0	& 0.5	& 0	& B	& A	& A	& B\\
    0	& 0.5	& 0.5	& B	& A	& B	& B\\
    0	& 0.75	& 0.25	& B	& B	& A	& A\\
    0	& 0.75	& 0.75	& B	& A	& B	& B\\
    0.25	& 0	& 0.25	& B	& B	& B	& B\\
    0.25	& 0	& 0.75	& B	& B	& B	& A\\
    0.25	& 0.25	& 0	& B	& B	& B	& B\\
    0.25	& 0.25	& 0.5	& A	& B	& A	& B\\
    0.25	& 0.5	& 0.25	& A	& A	& B	& B\\
    0.25	& 0.5	& 0.75	& B	& B	& B	& A\\
    0.25	& 0.75	& 0	& B	& B	& B	& B\\
    0.25	& 0.75	& 0.5	& B	& B	& A	& B\\
    0.5	& 0	& 0	& B	& B	& A	& A\\
    0.5	& 0	& 0.5	& B	& B	& A	& A\\
    0.5	& 0.25	& 0.25	& B	& B	& B	& A\\
    0.5	& 0.25	& 0.75	& B	& B	& B	& B\\
    0.5	& 0.5	& 0	& B	& B	& B	& B\\
    0.5	& 0.5	& 0.5	& B	& B	& B	& B\\
    0.5	& 0.75	& 0.25	& B	& A	& A	& B\\
    0.5	& 0.75	& 0.75	& B	& B	& B	& A\\
    0.75	& 0	& 0.25	& B	& A	& B	& B\\
    0.75	& 0	& 0.75	& A	& B	& A	& A\\
    0.75	& 0.25	& 0	& B	& B	& A	& A\\
    0.75	& 0.25	& 0.5	& B	& A	& B	& A\\
    0.75	& 0.5	& 0.25	& B	& A	& A	& B\\
    0.75	& 0.5	& 0.75	& B	& B	& B	& A\\
    0.75	& 0.75	& 0	& B	& B	& A	& A\\
    0.75	& 0.75	& 0.5	& B	& B	& A	& A
  \end{tabular}
  \end{ruledtabular}
\end{table}

\begin{table}
  \caption{Directionally resolved SRO parameters calculated according to Eq.~\ref{eq:doSRO}.}\label{tab:doSRO}
  \begin{ruledtabular}
  \begin{tabular}{ccccc}
  & \multicolumn{4}{c}{mole fraction, $x_A$}\\
  &	0.125&	0.25&	0.375&	0.5\\\hline
  $\alpha_{1,x}$&	$0.000$&	$0.000$&	$0.000$&	$0.000$\\
  $\alpha_{1,y}$&	$0.000$&	$0.000$&	$0.000$&	$0.000$\\
  $\alpha_{1,z}$&	$0.000$&	$0.000$&	$0.000$&	$0.000$\smallskip\\
  $\alpha_{2,x}$&	$-0.143$&	$0.000$&	$-0.067$&	$0.000$\\
  $\alpha_{2,y}$&	$-0.143$&	$0.000$&	$-0.067$&	$0.000$\\
  $\alpha_{2,z}$&	$-0.143$&	$0.000$&	$-0.067$&	$0.000$\smallskip\\
  $\alpha_{3,x}$&	$0.000$&	$-0.083$&	$-0.133$&	$0.000$\\
  $\alpha_{3,y}$&	$-0.143$&	$0.000$&	$-0.133$&	$0.000$\\
  $\alpha_{3,z}$&	$0.000$&	$0.000$&	$0.067$&	$0.000$\smallskip\\
  $\alpha_{4,x}$&	$-0.143$&	$0.000$&	$-0.067$&	$0.000$\\
  $\alpha_{4,y}$&	$-0.143$&	$-0.167$&	$-0.067$&	$0.000$\\
  $\alpha_{4,z}$&	$-0.143$&	$-0.167$&	$-0.067$&	$0.000$\smallskip\\
  $\alpha_{5,x}$&	$0.000$&	$0.000$&	$0.000$&	$0.000$\\
  $\alpha_{5,y}$&	$0.000$&	$0.000$&	$0.000$&	$0.000$\\
  $\alpha_{5,z}$&	$0.000$&	$0.000$&	$0.000$&	$0.000$
  \end{tabular}
  \end{ruledtabular}
\end{table}

\end{document}